\documentclass[sort&compress,5p]{elsarticle}
\usepackage{natbib}
\usepackage{siunitx}

\begin{document}
\begin{frontmatter}
\title{Physics with Radioactive Ion Beams at the Nuclear Science Laboratory}
\author[ND,cor1]{W.S. Porter}
\author[ND]{D.W. Bardayan}
\author[ND]{M. Brodeur}
\author[ND]{C. Dembski}
\author[ND]{P.D. O'Malley}
\author[ND]{F. Rivero}
\author[ND]{W.W. von Seeger}
\author[ND]{C. Boomershine}
\author[ND]{O. Bruce}
\author[ANL,ND]{S.R. Carmichael}
\author[ND,FRIB,MSU]{S. Coil}
\author[ND,ANL]{A.M. Houff}
\author[ND]{J.J. Kolata}
\author[ND]{D. Schroeder}
\author[ANL]{A.A. Valverde}
\author[ND]{R. Zite}


\address[ND]{Department of Physics \& Astronomy, University of Notre Dame, Notre Dame, IN 46556, USA}
\address[ANL]{Physics Division, Argonne National Laboratory, Lemont, Illinois 60439, USA}
\address[FRIB]{Facility for Rare Isotope Beams, Michigan State University, East Lansing, Michigan 48824, USA}
\address[MSU]{Department of Physics and Astronomy, Michigan State University, East Lansing, Michigan 48824, USA}

\cortext[cor1]{Corresponding author: wporter@nd.edu}

\begin{abstract}
The University of Notre Dame is home to a rich history of nuclear science efforts, including some of the earliest developments in the production and creation of radioactive ion beams (RIBs). The resulting \textit{TwinSol} facility stands as the longest continuously-operating RIB facility in the United States, and has enabled a wide range of experimental efforts in nuclear structure, astrophysics and fundamental symmetries through in-flight RIBs one to two nucleons from stability. A recent upgrade to \textit{TwinSol} via a third solenoid constituted the founding of the \textit{TriSol} facility, which provides cleaner, better-focused RIBs to downstream experiments. In this paper, we present a summary of RIB science at the Nuclear Science Laboratory (NSL) with both \textit{TwinSol} and \textit{TriSol} and discuss recent upgrades and future experimental plans at the facilities.
\end{abstract}
\end{frontmatter}

\section{Introduction}

The University of Notre Dame boasts a long history of nuclear science research dating back to the 1930s, including the production of some of the earliest radioactive ion beams (RIBs) at a university-based facility \cite{Becchetti1989}. A strong foothold in the realm of RIB science came in the late 1990s with the first operation of \textit{TwinSol} \cite{Lee1997}, a two solenoid separator system that enables the creation of a secondary beam via separation of reaction products from the incoming primary beam. This system enables production of high-intensity, low-contamination RIBs at $\sim 1 - 5$ MeV/u from $A = 3 - 40$. Historically and currently, these capabilities enable studies of fundamental symmetries, nuclear structure and reactions, and nuclear astrophysics with nuclei a few nucleons from stability.

The advantage in separation capability offered by a two solenoid system can be improved by the addition of a third solenoid, which enables better separation of unwanted secondary products from the secondary beam of interest. This has been accomplished with the addition of a superconducting 5T magnet after a dipole switching magnet, extending \textit{TwinSol} into \textit{TriSol} \cite{OMalley2023}. The improved secondary beam quality, alongside new access to RIBs previously challenging to produce cleanly, enables a new swath of experiments.

In this publication, we will discuss the status and ongoing science program of both \textit{TwinSol} and \textit{TriSol}. Section \ref{sec:TwinSol} discusses \textit{TwinSol}, while Section \ref{sec:TriSolup} focuses on the upgrades and improvements with \textit{TriSol}. Section \ref{sec:TriSolprogram} discusses the ongoing and planned science program with \textit{TriSol}.

\section{\textit{TwinSol}}
\label{sec:TwinSol}

Though RIB production at the NSL has a more than 35 year history, the foundation of the \textit{TwinSol} facility in 1997 established a new era of radioactive beam science at university accelerator facilities. Consisting of two 6 T, 30 cm bore solenoid magnets, \textit{TwinSol} enables in-flight separation of isotope beams based on their magnetic rigidity. These solenoids are positioned directly after a gas target, traditionally filled with deuterium or $^3$He, downstream of the NSL's FN 10MV tandem accelerator for creation and separation of 1-5 MeV/u RIBs a few nucleons from stability. \textit{TwinSol} constitutes the longest continuously-operating RIB facility in the US, and has enabled a myriad of experiments. It is still under operation today, providing RIBs to the newly-installed third solenoid as well as the Superallowed Transition Beta-Neutrino Decay Ion Coincidence Trap (St. Benedict) facility, which aims to measure the beta-neutrino angular correlation parameter of various mixed mirror beta decays for precision tests of the Standard Model \cite{Porter2023}. Further details on this facility can be found in the associated publication by Zite et al.

The extension of the downstream beamline of \textit{TwinSol} into an adjacent area through a meter-thick shielding wall provided necessary space and flexibility for larger experimental setups, including the aformentioned St. Benedict facility, as well as reducing $\gamma$ ray and neutron background. The beam size and quality of RIBs deliverable by \textit{TwinSol} at this new location further downstream hampered many experiments, and thus the addition of a new solenoid was needed for improved beam optics.

\section{Upgrades and the \textit{TriSol} facility}
\label{sec:TriSolup}

\begin{figure}
\includegraphics[width=\columnwidth]{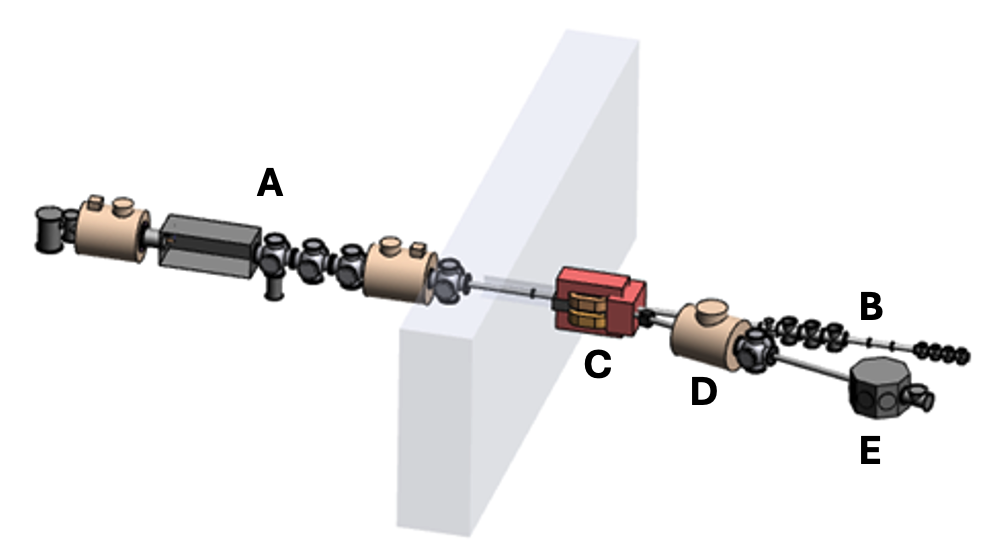}
\caption{A schematic of the \textit{TwinSol} and \textit{TriSol} facilities at the NSL. The leftmost two solenoids (A) in cream constitute \textit{TwinSol}, which now primarily delivers beams to St. Benedict (B). The rightmost solenoid in cream (D) established the \textit{TriSol} beamline, which receives beam from upstream via the dipole switching magnet labeled in red (C). Downstream, a representative octagonal scattering chamber (E) is shown. \label{fig:TriSol}}
\end{figure}

In 2021, a 1m-long, 5T magnet was installed downstream of a thick shielding wall to provide higher-quality RIBs and a space for minimized neutron and gamma ray background for experiments using such RIBs. This third solenoid is located on the 15$^\circ$ line directly after a dipole switching magnet inherited from Oak Ridge National Laboratory, and establishes the \textit{TriSol} facility. A schematic of the facility relative to \textit{TwinSol} and St. Benedict is shown in Figure \ref{fig:TriSol}. Full commissioning of the facility was completed in 2022 using two ($^3$He,n) reactions to produce secondary $^8$B and $^{14}$O beams \cite{OMalley2023}. Beam spot size determination with a Si detector positioned after the third solenoid demonstrated a five times reduction in spot size of the $^8$B beam with an appropriate solenoidal field when compared with the solenoid off. This result agreed nicely with simulation and demonstrated successful improvement of secondary beam quality with \textit{TriSol}. Further optimization of beam spot size and divergence angle is possible with slightly different detector and/or experiment locations.

Tests with an $^{14}$O secondary beam demonstrated the ability for \textit{TriSol} to improve the purity of certain secondary beams via the suppression of scattered primary beam. With the addition of the third solenoid alongside a charge-stripping Mylar foil located at the crossover point between the first and second solenoids, the first solenoid is now able to be tuned to pass a lower charge state, while the second and third solenoids are tuned to pass the charge state of interest. The result is minimal impact on the $^{14}$O beam, which is changed from a 7$^+$ charge to 8$^+$ in between the first two solenoids and is well-tuned through the system as intended, but a large impact on the scattered primary $^{12}$C beam, which now is directed into a pair of crossover slits. The result is an improvement in the contaminant to beam of interest ratio by a factor of 4, an improvement that enables \textit{TriSol} to conduct a variety of experiments previously inaccessible with \textit{TwinSol} purity capabilities. Further details on commissioning can be found in \cite{OMalley2023}.


\section{Experimental Program with \textit{TriSol}}
\label{sec:TriSolprogram}

Since its commissioning, \textit{TriSol} has been used in a variety of different experiments of interest to nuclear structure, astrophysics and fundamental symmetries. In this section, the status and ongoing work of many of these experiments and areas, as well future measurement plans and capabilities, are presented.

\subsection{Scattering Experiments}

\begin{figure}
\includegraphics[width=\columnwidth]{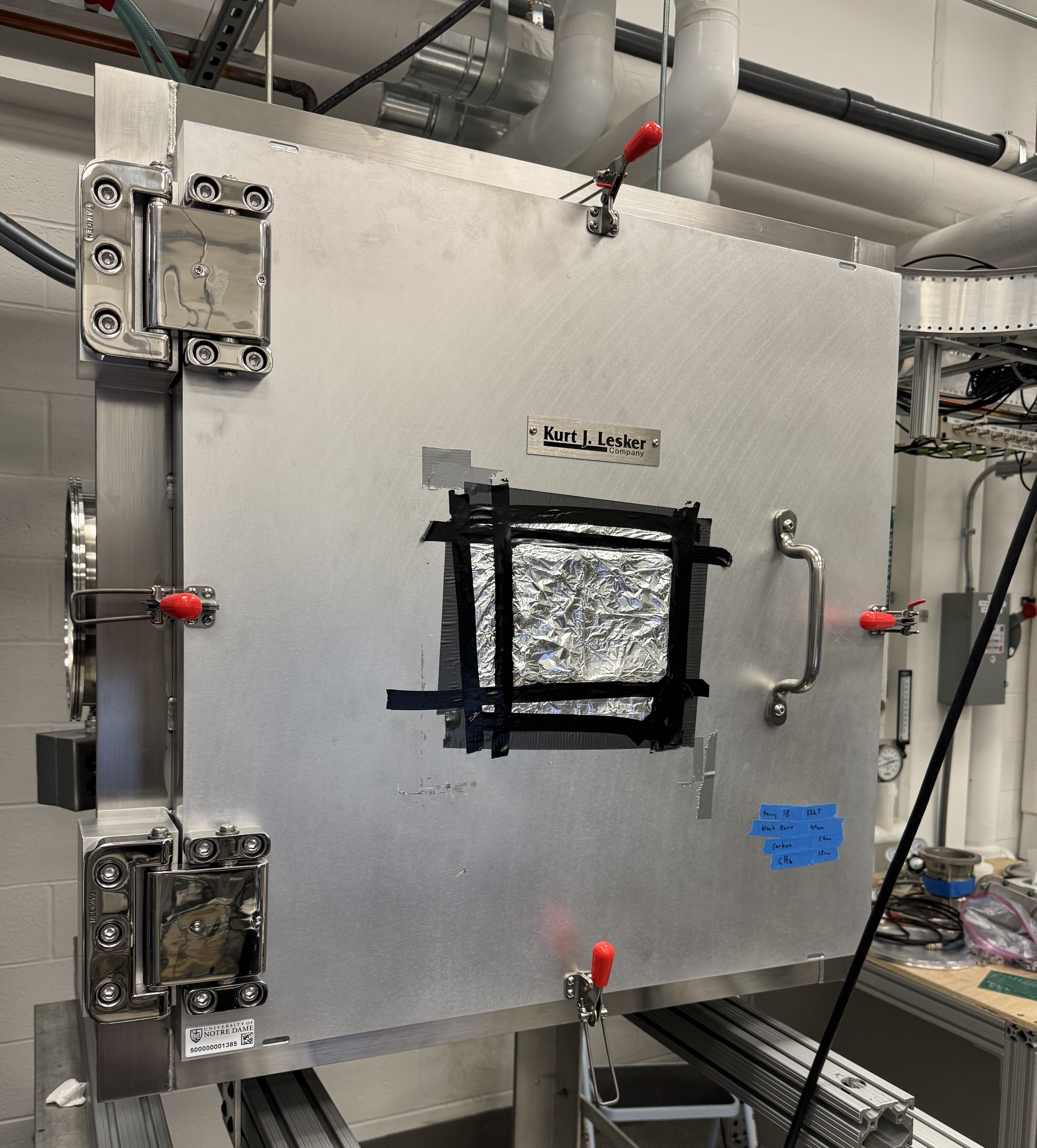}
\caption{An image of the new scattering chamber for the \textit{TriSol} facility. \label{fig:new_scatt}}
\end{figure}

Many of the inaugural experiments utilizing the \textit{TriSol} facility involved a target and/or detector system mounted in a general purpose scattering chamber positioned after the third solenoid. These experiments include studies of r-process seed nucleus production via the Coulomb breakup of $^6$He on $^{208}$Pb \cite{Godos2024}, scattering measurements at sub-Coulomb barrier energies with $^8$B and $^7$Be \cite{Palli2023,Palli2024}, and indirect studies of the $^{14}$O($\alpha,p$) reaction via the $^{17}$F($p,p'$) reaction for Type-1 x-ray bursts \cite{Coil2025}. These experiments all used an older scattering chamber, which was limited in its accessibility and functionality.

Recently, a new scattering chamber has been designed and fabricated, and offers a significant upgrade over the previously-used chamber. The large door provides improved access for constructing setups, and multiple ports along the beam axis enable increased flexibility in target location, where the old chamber was limited to precarious access through its topside and a central target location. Additionally, the base vacuum was limited to \num{3E-5} Torr when pumped with two 300 L/s turbomolecular pumps; the new chamber achieves a \num{8E-7} Torr base pressure with only one of these turbos. A picture of this chamber is shown in Figure \ref{fig:new_scatt}. This chamber completely replaces the one previously used, which has since been decommissioned, and will feature in several planned \textit{TriSol} experimental campaigns in the near future.

\subsection{Half-Lives}

\begin{figure}
\includegraphics[width=\columnwidth]{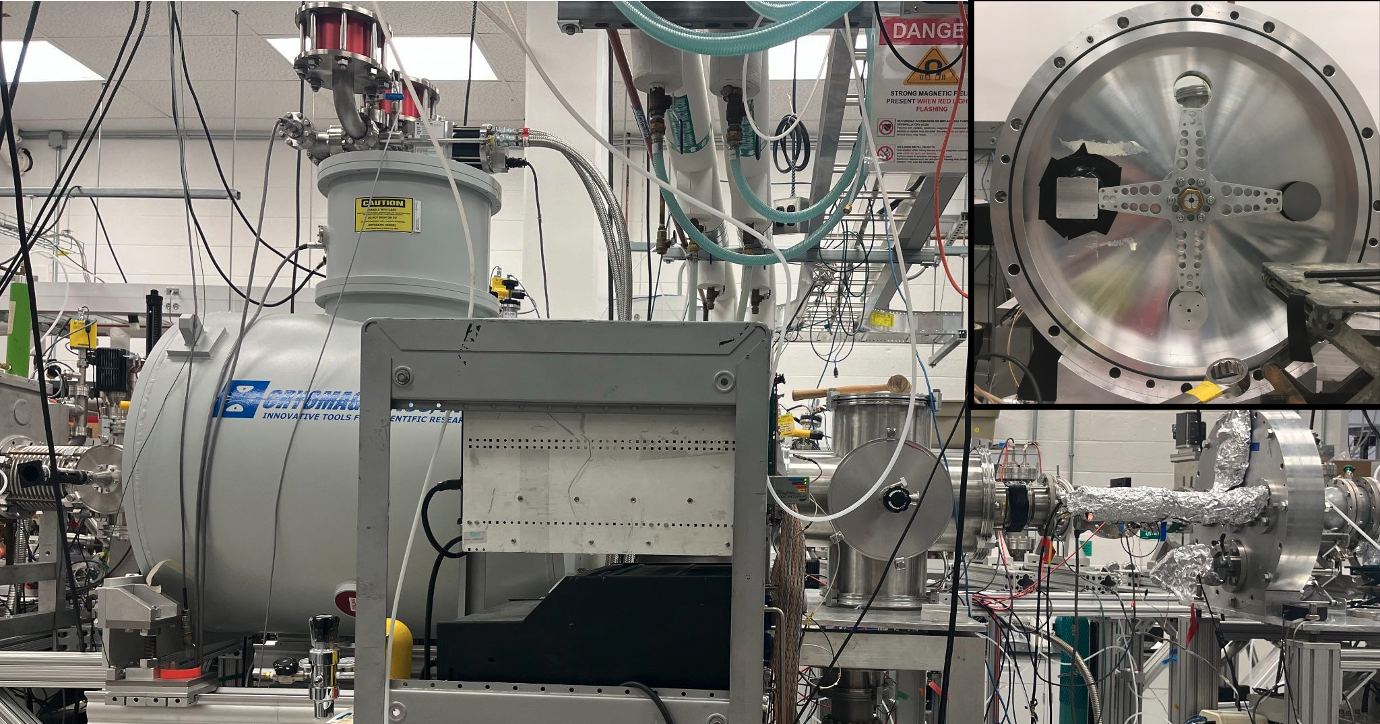}
\caption{An image of the beta counter system as installed at the \textit{TriSol} facility. The top-right corner inset image shows the rotation arms of the counter that move the Ta foil in and out of the path of the beam and scintillator system \label{fig:t12_setup}.}
\end{figure}

The NSL has been home to a thriving precision $\beta$ decay half-life measurement program for nearly a decade, including many mixed mirror decays of interest for measurement via the St. Benedict facility \cite{Long2022}. Measurements of $\beta$ decays are made by implanting the mother RIB of interest on a thick, high-purity Au foil, which is subsequently rotated 180 degrees directly in front of a 1mm-thick plastic scintillator backed by a photomultiplier tube (PMT) for $\beta$ detection. An image of the detector system employed can be seen in Figure \ref{fig:t12_setup}.

\begin{figure}
\centering
\includegraphics[width=0.65\columnwidth]{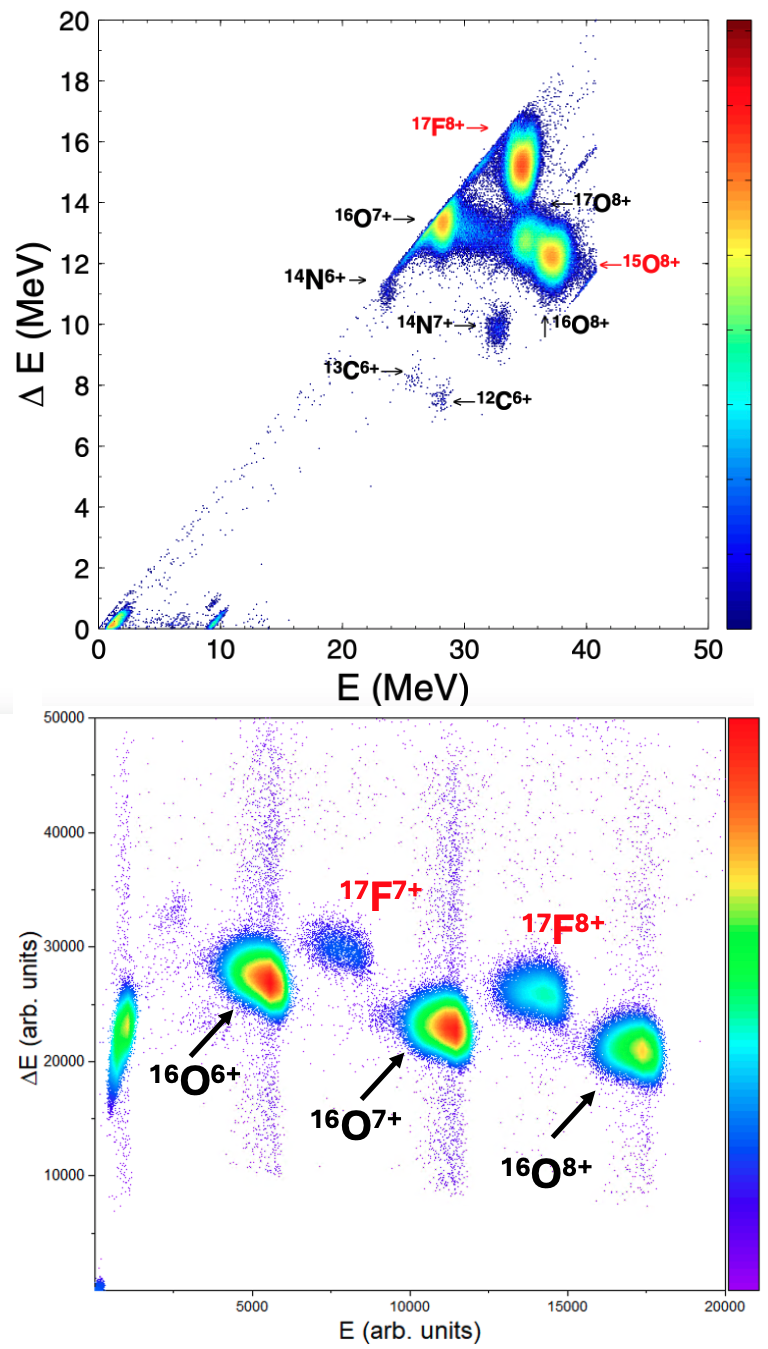}
\caption{PID plots comparing the production of $^{17}$F with \textit{TwinSol} (top) versus with \textit{TriSol} (bottom). A clear reduction in contaminants, most notably of the radioactive contaminant $^{15}$O, is apparent when using the \textit{TriSol} facility. \label{fig:17F_ACKBAR}.}
\end{figure}

Historically, measurements were made using RIBs produced from \textit{TwinSol}, including a range of isotopes from $^{11}$C to $^{29}$P. Limitations in separation capabilities of the two-solenoid system prevented many challenging measurements, including heavier mirror isotopes of interest. The advent of the \textit{TriSol} facility enables significantly improved measurements, as demonstrated in Figure \ref{fig:17F_ACKBAR} taken during a recent $^{17}$F half-life measurement campaign. The significant reduction in presence of radioactive contaminants, primarily indicated by a decrease in $^{15}$O as detected by an $E-\Delta E$ telescope, suggests marked improvement of $^{17}$F RIB purity with the \textit{TriSol} facility. Alongside the integration of a Mesytec digital DAQ to the half-life measurement system, this enables both precision improvements on previously measured isotopes (e.g. $^{17}$F, $^{29}$P) and access to new heavy isotopes previously unmeasured at the NSL (e.g. $^{31}$S, $^{41}$Sc).

\subsection{ATHENA}

\begin{figure}
\includegraphics[width=\columnwidth]{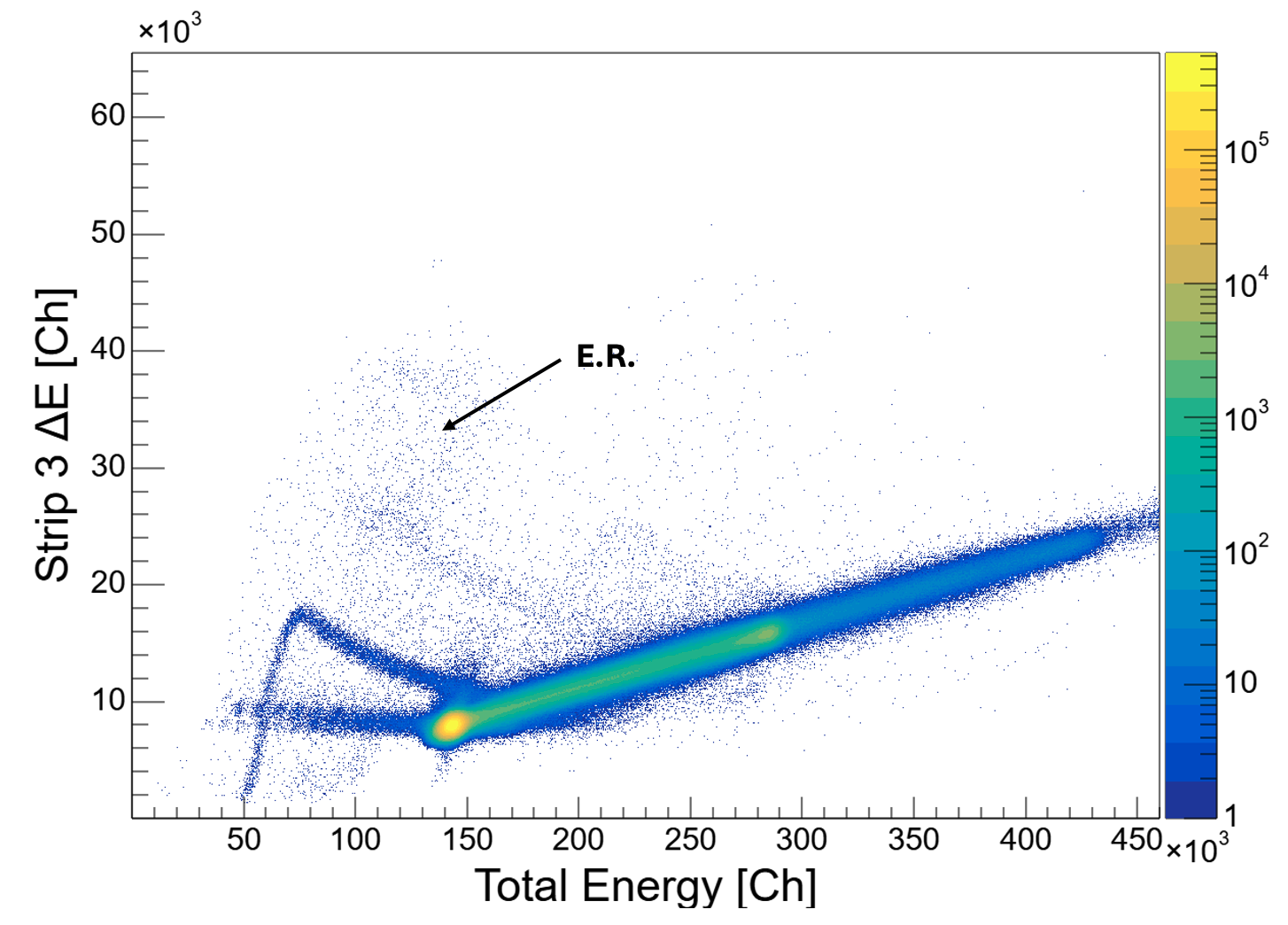}
\caption{A plot of detected energies in a strip of ATHENA during commissioning with $^{12}$C+$^{12}$C fusion taken a counting rate of 100 kHz. The presence of different beam-like, scattering and evaporation residue (E.R.) events similar to that in \cite{Blankstein2023} indicates success measurement at a counting rate previously inaccessible. \label{fig:100kHz_ATHENA}}
\end{figure}

Many of the most important low to medium-mass astrophysical reactions are helium-induced reactions that have traditionally been challenging to measure directly. The advent of active target detectors, in particular Multi-Sampling Ionization Chamber (MuSIC) detectors, has enabled measurements of ($\alpha,n$) and ($\alpha,p$) total cross-sections via detection of heavy recoils in a gas-filled ionization chamber.

The Active Target High Efficiency Detector for Nuclear Astrophysics (ATHENA) \cite{Blankstein2023} at the NSL aims to measure a variety of $\alpha$-induced reactions, including the $^{14}$O($\alpha,p$)$^{17}$F reaction relevant for processes fueling Type-1a X-ray bursts, and the $^{15}$O($\alpha,p$)$^{18}$F reaction for understanding CNO cycle pathways in novae. Measurements of these reaction total cross-sections were previously untractable given a limited counting rate of 5 kHz, due in part to a high DAQ dead time and low-quality of pure He as a counting gas. The recent implementation of a Mesytec digital DAQ system, higher cathode voltages for improved recoil detection resolution and a better counting gas via He doped with CF$_4$ allows for the higher counting rates needed to make measurements feasible. Offline commissioning of these ATHENA upgrades with $^{12}$C+$^{12}$C fusion indicates that counts rates up to 150 kHz are possible, as shown in the characteristic strip plot in Figure \ref{fig:100kHz_ATHENA}.

\subsection{SSNAPD}


The solenoidal spectrometer concept, which employs an array of  silicon detectors placed along the magnetic axis of a large-bore solenoid to convert angular distribution measurements into position measurements free of kinematic compression, has been primarily leveraged in the field of direct reaction studies. The unique triple-solenoid configuration of the TriSol system will allow for a new use of this detector concept, leveraging the robust particle identification provided by the cyclotron period to provide high-precision measurements of charged particle decay branching ratios for nuclear states that influence astrophysically-relevant resonant reactions.

\begin{figure}
\includegraphics[width=\columnwidth]{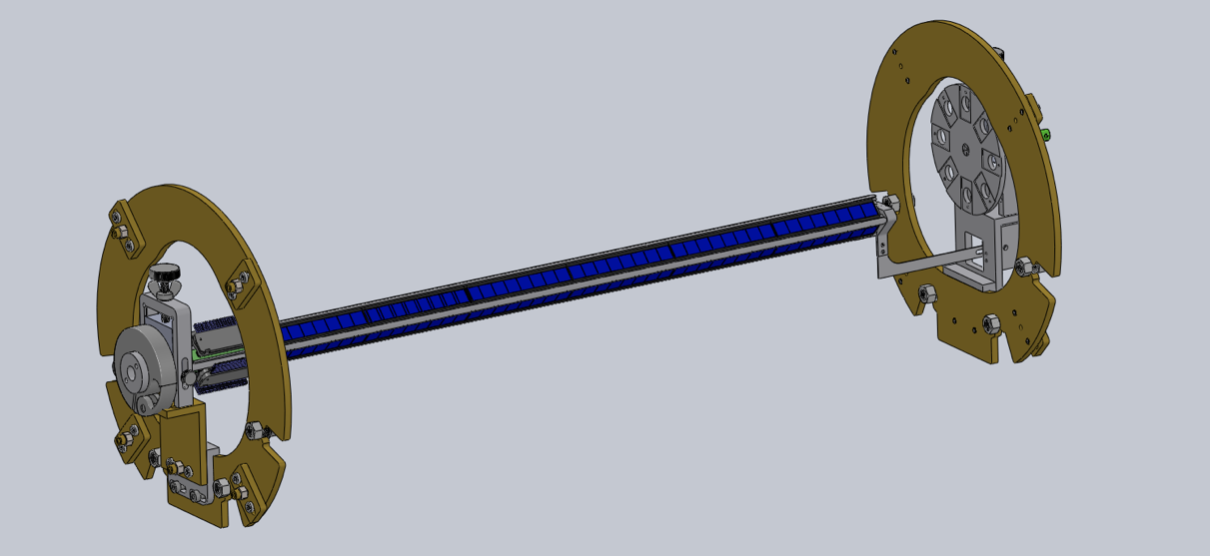}
\caption{A CAD drawing of the primary SSNAPD detector array that will be inserted on beam axis inside of the first solenoid of \textit{TriSol}. \label{fig:SSNAPD}}
\end{figure}

The Solenoid Spectrometer for Nuclear AstroPhysics and Decays (SSNAPD) converts the first TriSol solenoid into a traditional solenoidal spectrometer array, shown in Figure \ref{fig:SSNAPD}, covering backwards laboratory angles from the downstream target. The second and third solenoids act as a magnetic spectrometer to separate reaction ejectiles produced in e.g. transfer reactions used to populate the excited states of interest. Coincidences between reaction ejectiles in the downstream $\Delta E-E$ telescope and decay particles in the solenoidal spectrometer enable a direct determination of charged particle branching ratios. 

In particular, studies of $^{19}$Ne$^*$ $\alpha$-branching for the critical $^{15}$O($\alpha,\gamma$) reaction via the $^{19}$F($^3$He,$t$) charge-exchange reaction are planned for SSNAPD. Commissioning of SSNAPD is underway, and further studies of other alpha, proton and deuteron reactions relevant for nuclear astrophysics and structure are planned.

\section{Conclusion}

The nuclear physics program at the NSL boasts many experiments utilizing a variety of RIBs historically produced in-flight using the \textit{TwinSol} facility. The advent of the \textit{TriSol} facility present new opportunities in fundamental symmetries, nuclear structure and astrophysics research, employing an array of experimental setups and detectors, ultimately positioning the NSL to play an integral role in these research areas for the forseeable future.

\section*{Acknowledgments}
This work is supported in part by the National Science Foundation under Grant Nos. PHY-2310059 and PHY-2117687, and by the University of Notre Dame.


\end{document}